\documentclass[a4paper,11pt]{article}
\usepackage{jcappub} 
\usepackage{lineno} 
\usepackage{appendix}
\usepackage{orcidlink}
\usepackage{hyperref}

\newcommand{\appropto}{\mathrel{
    \vcenter{
        \offinterlineskip
        \halign{
            \hfil$##$\cr\propto\cr\noalign{\kern1pt}\sim\cr\noalign{\kern-1pt}
        }
    }
}}

\makeatletter
\gdef\@fpheader{}
\makeatother

\arxivnumber{2405.11628}
\title{\boldmath Generalised conditions for rapid-turn inflation}

\author[a,e]{Raúl~Wolters\,\orcidlink{0009-0006-9928-6214},}
\author[b,c]{Oksana~Iarygina\,\orcidlink{0000-0002-6222-1664}}
\author[a,d]{and~Ana~Achúcarro\,\orcidlink{0000-0002-2325-9678}}

\affiliation[a]{Institute Lorentz of Theoretical Physics, Leiden University, 2333 CA Leiden, The Netherlands}
\affiliation[b]{Nordita, KTH Royal Institute of Technology and Stockholm University, Hannes Alfv\'ens v\"ag 12, 10691 Stockholm, Sweden}
\affiliation[c]{The Oskar Klein Centre, Department of Astronomy, Stockholm University, 10691 Stockholm, Sweden}
\affiliation[d]{ Department of Theoretical Physics, University of the Basque Country, UPV-EHU 48080 Bilbao, Spain}
\affiliation[e]{Institute for Theoretical Physics, Utrecht University, 3584 CE Utrecht, The Netherlands}

\emailAdd{r.wolters@uu.nl, oksana.iarygina@su.se, achucar@lorentz.leidenuniv.nl}

\abstract{Rapid-turn slow-roll inflationary trajectories have been shown to be an attractor in two-field models, provided the turn rate is near constant and larger than the slow-roll parameters. These trajectories can produce primordial spectra consistent with current observations on CMB scales. 
We present the generalized consistency condition for sustained rapid-turn inflationary trajectory with two fields, arbitrary field-space metric and potential valid for any value of the turn rate. This has to be supplemented by a second condition to ensure slow roll evolution. Both conditions together  constitute a tool 
 to identify inflationary trajectories with arbitrary values of the turning rate without having to solve the equations of motion. 
 We present a Python package for 
 the numerical identification of regions in field-space and parameter space that allow for rapid-turn trajectories.  }

\begin{document}
\maketitle
\flushbottom
\section{Introduction}
Inflation is the leading framework for the physics of the very early universe. It provides an elegant solution for the origin of the primordial perturbations that seed all structure in the universe, which is in excellent agreement with the latest observational tests. Typically, inflation is described by a single scalar field, slowly rolling down its potential. However, models of particle physics beyond the Standard Model motivate many scalar fields and their presence in the inflationary epoch may significantly influence cosmological predictions (see \cite{Achucarro:2022qrl, Cicoli:2023opf,Christodoulidis:2023eiw,Ellis:2023wic} for recent reviews).  

Observations of the Cosmic Microwave Background (CMB) \cite{Planck:2018jri} highly constrain the power spectrum of primordial fluctuations, which is very close to scale-invariant on the largest scales. This restricts the inflaton velocity and its fractional rate of change to be small compared to the Hubble rate.
In single field inflation, these slow-roll conditions translate into the requirement for a sufficiently flat scalar potential, $V$.  By contrast,  in multi-field models, the slow roll conditions only imply flatness of the potential {\em  along the trajectory}, and are compatible with steep potentials provided the turn rate of the inflationary trajectory is large enough \cite{Achucarro:2010da, Peterson:2010np, Yang:2012bs,Hetz:2016ics,Achucarro:2018vey,Aragam:2019omo}. An important question is whether this non-geodesic motion can be sustained for enough e-folds of expansion to generate primordial spectra compatible with the current observations.

Rapid-turn models of inflation are particularly interesting in view of upcoming high-precision observational experiments of CMB and the Large Scale Structure (LSS). Unlike single-field models of inflation that produce nearly Gaussian fluctuations with a very small amplitude of the bispectrum, suppressed by the slow-roll parameters \cite{Maldacena:2002vr,Creminelli:2004yq}, multi-field rapid-turn models of inflation may generate potentially detectable values of non-Gaussianity \cite{Garcia-Saenz:2018vqf,Fumagalli:2019noh,Bjorkmo:2019qno,Ferreira:2020qkf,Garcia-Saenz:2019njm, Iarygina:2023msy}.

In this paper we are primarily interested in 
rapid-turn models consistent with slow roll. In the case of two-field inflation, Bjorkmo \cite{Bjorkmo:2019fls} showed the existence of a “slow-roll rapid-turn" attractor (SRRT) — for which the turn rate $\omega$ (in e-folds) is approximately constant and $\omega^2 \gg {\cal O}(\rm slow-roll)$ —  which provides a unified description for several known examples in the literature \cite{Achucarro:2012sm, Achucarro:2012yr,Brown:2017osf,
Mizuno:2017idt,Christodoulidis:2018qdw, Garcia-Saenz:2018ifx, Bjorkmo:2019aev,Achucarro:2019pux,Achucarro:2019mea,Chakraborty:2019dfh,Aragam:2020uqi,Aragam:2021scu,Renaux-Petel:2021yxh}. 
We will adopt this terminology here\footnote{Note that the words “rapid turn” are sometimes used in the literature to mean $\omega >1$, but also for sudden turns where $\omega$ changes rapidly, regardless of its size. In this paper we restrict ourselves to the case where $\omega$ is slowly varying $d\omega / dN \ll \omega$ and make no other assumptions about its size other than  $\omega^2 \gg {\cal O}(\rm slow-roll)$.}. A key ingredient in the analysis of \cite{Bjorkmo:2019fls,Bjorkmo:2019aev} is the use of a vector basis in field space \cite{Gao:2012uq} that is associated with the potential and does not require information about the specific trajectory (see
\cite{Christodoulidis:2022vww, Christodoulidis:2019jsx,Christodoulidis:2019mkj}
for an alternative approach). 

 Several authors have extended these results,  most notably ref \cite{Aragam:2020uqi}. The situation with more than two fields is considerably more complicated, see e.g 
\cite{Aragam:2019omo,Bjorkmo:2019aev,Aragam:2020uqi,Christodoulidis:2021vye,Christodoulidis:2022vww,Christodoulidis:2023eiw,
Pinol:2020kvw,Achucarro:2018ngj} for partial results. 
 Reference 
 \cite{Aragam:2021scu} 
 explored the connection to supergravity and surveyed some multifield supergravity models in the literature, and concluded SRRT trajectories are rare. 
 It would be very useful to have a  tool to pinpoint the location of these slow-roll-rapid-turn attractors in field space, if they exist, for any given multifield model. This paper is a first step in that direction, for the case of two fields (see also \cite{Aragam:2020uqi} for a slightly different implementation that extends to the case of more than two fields).

Recently, the authors of \cite{Anguelova:2022foz, Anguelova:2023mpe} derived a consistency condition for a two-field potential to support SRRT in the case where $\omega\gg 1 $ and
the trajectory is almost orthogonal to $\nabla V$. As we will show, this assumption is too restrictive to capture even some of the known examples of rapid-turn slow-roll inflation in the literature\footnote{A related problem is that, if the trajectory becomes completely horizontal (orthogonal to grad $V$),  inflation enters a (transient) ultra-slow-roll regime \cite{Inoue:2001zt, Kinney:2005vj} which violates the slow-roll conditions and gives very different phenomenology. This is beyond the scope of this paper. We focus on sustainable SRRT  trajectories, that continuously roll ``downhill" in V.}.  Here we generalize the analysis in \cite{Anguelova:2022foz} to obtain a consistency condition that is valid for all values of the turning rate and all inclinations of the trajectory, and show that it reproduces their result in the right limit. Further, we show that it does not guarantee slow roll and a second condition is needed to identify SRRT trajectories.

To summarize, in this work we present two \textit{consistency conditions} that allow us to identify slow-roll-rapid-turn trajectories in two-field inflation without any assumptions about the magnitude of the turning rate other than $\omega^2\gg {\cal O}(\rm slow-roll)$ :
\begin{eqnarray*}
    &\frac{V_{ww}}{V}=3+3\left(\frac{V_{vv}}{V_{vw}}\right)^2+\frac{V_{vv}}{V}\left(\frac{V_{vw}}{V_{vv}}\right)^2 \\
&
\frac{3\varepsilon_V}{\varepsilon_V+
\left( \frac{V_{ww}}{V} - \frac{V_{vw}^2}{V V_{vv}} \right) } \ll 1 
\end{eqnarray*}
where $\varepsilon_V=\frac{1}{2}\frac{\nabla_aV\nabla^aV}{V^2}$ is the usual, multifield potential slow-roll parameter and the second condition is needed to ensure slow roll (this can be done in a number of ways). In these expressions, $V_{\alpha\beta}=\alpha^a\beta^b\nabla_a\nabla_bV(\phi)$
are the components of the (covariant) Hesse matrix of the potential $V$ projected along the unit vectors $\alpha, \beta$ = $\{ v,w \}$. The vectors of interest are $v$, the unit vector parallel to the gradient of $V$, and $w$, a unit vector perpendicular to $v$. The first equation is a rapid-turn condition that generalizes the condition from \cite{Anguelova:2022foz} to steeper trajectories (smaller turn rates). Slow-roll is not guaranteed by the rapid-turn condition so the second inequality imposes slow roll \textit{on the rapid-turn trajectory} (and nowhere else).
The consistency conditions do not require any information about the trajectory, only about the scalar potential and its derivatives, and about the field space metric.

We numerically implement the consistency conditions and present the software package written in Python, \texttt{inflatox}, that for a given two-field potential and a field-space metric allows one to identify regions in field space
(and in parameter space) where the consistency conditions are satisfied. This allows for easy tracking of the existence of (unknown) rapid-turn two-field inflationary attractors for a given model. To check the accuracy of the consistency conditions we compare the results with some known examples of rapid-turn inflationary trajectories in the literature. We show that the new generalized rapid-turn condition correctly identifies rapid-turn trajectories and we provide a comparison with the condition from \cite{Anguelova:2022foz}. We demonstrate that the rapid turn condition does not guarantee
slow-roll inflation by itself and show that adding the second condition selects the correct SRRT trajectories.

The structure of the paper is as follows. In Section \ref{sec:background} we introduce our notation and the background equations of motion together with the slow-roll and turning parameters relevant to the paper. In Section \ref{sec:cc} we present the derivation of the slow-roll rapid-turn consistency conditions. We further discuss the approximations taken in the transformation from kinematical to potential basis \footnote{   An exact expression for the angle between the kinematical and potential bases is given in Appendix \ref{appendix:delta}.} and derive a more general rapid-turn condition valid for arbitrary values of the turn rate. 
  In Section \ref{sec:numerics} we
introduce the \texttt{inflatox} package and discuss the numerical implementation of the consistency conditions.
We illustrate our findings with three different models of rapid-turn inflation. We conclude in Section  \ref{sec:conclusion}.

\section{Background Dynamics} \label{sec:background}

In this work we study two-field models of inflation minimally coupled to gravity, described by the action 
\begin{equation}
    \label{eq:multifield_action}
    S=\int d^4x\sqrt{-g}\left[\frac{M_{\rm pl}^2}{2}R-\frac{1}{2}g^{\mu\nu}G_{ab}(\phi)\partial_{\mu}\phi^a\partial_{\nu}\phi^b-V(\phi^a)\right],
\end{equation}
where $g_{\mu\nu}$ 
is the spacetime metric, $R$ is the Ricci scalar of $g_{\mu\nu}$, 
$G_{ab}(\phi)$ is the field-space metric and $V(\phi^a)$ is the multi-field potential. We use Latin indices to denote scalar fields and Greek indices for spacetime
coordinates. Throughout the paper we work in natural units and set $M_{\rm pl}=1$. We use the Friedmann-Robertson-Lemaître–Walker (FRLW) metric
\begin{equation}
ds^2=-dt^2+a(t)^2 \delta_{ij}dx^i dx^j,
\end{equation}
where $i,j$ indicate the spatial directions.
For the action \eqref{eq:multifield_action} the equation of motion for homogeneous fields $\phi^a(t)$ in FRLW metric is
\begin{equation}
\label{eq:multifield_eom}
        D_t\dot{\phi}^a+3H\dot{\phi}^a+\nabla^aV(\phi)=0,
\end{equation}
where $ D_t$ is the directional covariant derivative along a path parameterized
by the cosmic time $t$ and defined as $D_{t}=\dot{\phi}^a \nabla_a$, with $\nabla_a$ being the covariant derivative in field space. The directional covariant derivative $D_t$ acts on an arbitrary field-space vector $A^a=A^a(\phi)$ as
\begin{equation}
D_{t} A^a = \partial_t \phi^b \nabla_b A^a = \partial_t A^a+\Gamma^a_{bc}\dot{\phi}^bA^c,
\end{equation}
where $\Gamma^a_{bc}$ are the Christoffel symbols computed using $G_{ab}$.

Let us now define the slow-roll parameters that ensure the successful prolonged duration of inflation. \textit{The first slow-roll parameter}, $\varepsilon_H$, (also called the Hubble slow-roll parameter) parameterizes the deviation from the de Sitter expansion and is defined as
\begin{equation}
\label{eq:epsilon}
    \varepsilon_H=-\frac{\dot{H}}{H^2}=\frac{\frac{1}{2}G_{ab}\dot{\phi}^a\dot{\phi}^b}{ H^2}.
\end{equation}
Nearly exponential expansion during inflation happens when $\varepsilon_H \ll 1$. The prolonged duration of inflation requires $\varepsilon_H$ to be small for 
a sufficient number of e-folds. This is achieved when \textit{the second slow-roll parameter}, defined by
\begin{equation}
    \eta_H=\frac{\dot{\varepsilon}_H}{H \varepsilon_H}
\end{equation}
is small during inflation, i.e. $\eta_H \ll 1$ . Two conditions $\varepsilon_H \ll 1, \, \eta_H \ll 1$ are part of \textit{the slow-roll approximation}.

To further classify the background dynamics in the multi-field set-up let us define \textit{the covariant acceleration} of the field{\footnote{Note an opposite sign convention with Ref.~\cite{Anguelova:2022foz}. }}
\begin{equation}
    \eta^a=\frac{1}{H \dot{\varphi}}D_t \dot{\phi}^a,
\end{equation}
where $\varphi$ is the proper length along the inflationary trajectory, defined as 
\begin{equation}
    \dot{\varphi}=\sqrt{G_{ab}\dot{\phi}^a\dot{\phi}^b},
\end{equation}
which is determined up to constant translation. The field $\varphi$ plays the same role as the inflaton in single-field inflation. The Einstein equations in terms of $\varphi$ are 
\begin{equation}
    -2\dot{H}=G_{ab}\dot{\phi}^a\dot{\phi}^b=\dot{\varphi}^2, \quad 3H^2=V+\frac{1}{2}\dot{\varphi}^2.
\end{equation}

It is convenient to define a basis in the field space with basis vectors along and orthogonal to the inflationary trajectory. This is called \textit{the kinematical basis} and is convenient to describe perturbations, as one can immediately determine\footnote{See the recent discussion \cite{Cicoli:2021yhb} on the choice of entropy variables in multifield
inflation and references therein.}  adiabatic and isocurvature perturbations (along and orthogonal to the trajectory, respectively). The tangent and normal vectors are defined as\footnote{To avoid unphysical discontinuities and define $t^a$ and $n^a$ as continuous functions of time in the general multi-field case the normal vector can be defined as $n^a=s_n(t)\left( G_{bc}D_t t^b D_t t^c \right)^{-1/2}D_t t^a$, with $s_n(t)=\pm 1$. In the case of sustained rapid-turn inflation, the prefactor $s_N(t)$ can be omitted since $D_t t^a$ does not change direction. In the two-field case one can define $n_a=\sqrt{\det G}\, \epsilon_{ab}t^b$.  }
\footnote{It is worth noting the kinematical basis vectors are denoted as $(T,N)$ in \cite{Anguelova:2022foz}, with $(n, \tau)$ used for potential basis notation. We follow the notation from \cite{Bjorkmo:2019fls} for the basis vectors.}
\begin{equation}
    \label{eq:def_ta}
    t^a=\frac{\dot{\phi}^a}{ \dot{\varphi}}, \quad n^a=\left( G_{bc}D_t t^b D_t t^c \right)^{-1/2}D_t t^a ,
\end{equation}
with $t_a n^a=0$.
The covariant acceleration can be written in the kinematical basis as
\begin{equation}
    \eta^a=\eta_{\parallel}t^a+\omega n^a,
\end{equation}
where $\eta_{\parallel}$ is the covariant acceleration along the trajectory, also called \textit{the speed up rate}, since it quantifies the logarithmic rate of change of the field speed
\begin{equation}
       \label{eq:eta_parallel}
    \eta_{\parallel}=D_N\ln\dot{\varphi}=\frac{\ddot{\varphi}}{H\dot{\varphi}}=\frac{1}{2}\eta_H-\varepsilon_H.
\end{equation}
Here $D_N$ is the covariant derivative with respect to the number of e-folds, $N$, that is related to the cosmic time $t$ via $d/dN=H^{-1} d/dt$, where $H$ is the Hubble parameter. During slow-roll inflation $\varepsilon_H \ll 1, \, \eta_H \ll 1$, hence it follows $\eta_{\parallel}\ll 1$.
The projection of $\eta^a$ in the normal direction defines \textit{the turn rate}, $\omega$, which provides a dimensionless
measure of the deviation from a geodesic and is given by
\begin{equation}
    \omega=n_aD_Nt^a.
\end{equation}
Projecting the equations of motion \eqref{eq:multifield_eom} onto the kinematical basis $\{t^a,n^a\}$, we get
\begin{align}
\begin{split}
\label{eq:eom}
    \ddot{\varphi}+3H\dot{\varphi}+V_t&=0, \\
    \Omega\dot{\varphi}+V_n&=0,
    \end{split}
\end{align}
where $V_t=t^aV_a$ and $V_n=n^aV_a$ and
the parameter $\Omega$ is defined as
\begin{equation}
    \label{eq:def_omega}
    \Omega=n_aD_tt^a =\omega H.
\end{equation}
It is instructive to define
\textit{the potential slow-roll parameter} as 
\begin{equation}
    \label{eq:epsilonV}
    \varepsilon_V=\frac{1}{2}\frac{\nabla_aV\nabla^aV}{V^2}=\frac{1}{2}\frac{V_t^2+V_n^2}{V^2}.
\end{equation}
From the projection of the equation of motion in the normal direction \eqref{eq:eom}
one can find
\begin{equation}
\omega = -\frac{V_n}{H\dot\varphi} 
\qquad\qquad {\rm and} \qquad\qquad  
    \frac{1}{2}\frac{V_n^2}{V^2}=\frac{\varepsilon_H}{(3-\varepsilon_H)^2}\omega^2,
\end{equation}
where we used $V=H^2(3-\varepsilon_H)$ and the definition \eqref{eq:epsilon}.
Similarly, from the tangential projection
\begin{equation}\label{eq:VtviaepsH}
    \frac{1}{2}\frac{V_t^2}{V^2}=\frac{\varepsilon_H}{(3-\varepsilon_H)^2}\left(3+\eta_{\parallel}\right)^2,
\end{equation}
where we used again \eqref{eq:epsilon} and the equation (\ref{eq:eta_parallel}). Therefore, in two-field inflation one finds the following \textit{exact} relation between $\varepsilon_V$ and $\varepsilon_H$ 
\begin{equation}
    \label{eq:epsilon_V_exact}
    \varepsilon_V=\varepsilon_H\left[\left(\frac{3+\eta_{\parallel}}{3-\varepsilon_H}\right)^2+\left(\frac{\omega}{3-\varepsilon_H}\right)^2\right].
\end{equation}
In the regime where both $|\eta_{\parallel}|\ll1$ and $|\varepsilon_H|\ll1$ it simplifies to \cite{Hetz:2016ics}
\begin{equation}
    \label{eq:eV_eH}
    \varepsilon_V\simeq \varepsilon_H\left[1+\frac{\omega^2}{9}\right].
\end{equation}

\begin{figure}
    \centering
    \includegraphics[width=0.6\textwidth]{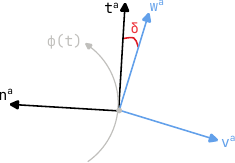}
    \caption{ Schematic drawing of a rapidly turning trajectory. The kinematical basis $\{t, n\}$ (tangent and normal to $\varphi$) and the potential basis $\{v, w\}$ (parallel and perpendicular to $\nabla V$) are drawn, with the angle $\delta=\angle(t,w)$ indicated.
    $\delta = 0$ corresponds to a trajectory with $V = $ const. Notice that the angle between $\nabla V$ and both $t$ and $n$ is always larger than $\pi/2$: during slow roll the trajectory must point ``downhill in V" to compensate for the Hubble friction,  and its turns are dictated by the centripetal force, opposite to $\nabla V$.}
    \label{fig:rapidturn}
\end{figure}

For general multi-field models of inflation, it might be challenging to find an analytic solution to equations of motion \eqref{eq:eom} due to the overall complexity of potentials and field-space manifolds. However, the choice of basis along and orthogonal to the gradient of potential can be extremely useful since the directions of basis vectors are known even before one has the full solution. \textit{The potential basis} is defined \cite{Bjorkmo:2019aev} via the unit vector along the gradient of the potential, $v^a$, and $w^a$, orthogonal to it 
\begin{equation}
    \label{eq:potential_basis}
    v^a=\frac{\nabla^aV}{\sqrt{G_{bc}\nabla^bV\nabla^cV}}=\frac{\nabla^aV}{|\nabla V|}, \quad w\perp v.
\end{equation}
At each point along the trajectory $\phi^a(t)$ one can construct a basis transformation matrix from the kinematical basis to the potential basis, as shown in figure \ref{fig:rapidturn}. Since $\{t, n\}$ and $\{v, w\}$ are both orthonormal bases, the basis transformation between the two is a simple rotation by an angle\footnote{This corresponds to the angle $\theta$ in \cite{Anguelova:2022foz}.} $(\pi/2 + \delta)$

\begin{equation}
    \label{eq:basistransformation}
    \begin{bmatrix}
        v\\
        w
    \end{bmatrix}=
        \begin{bmatrix}
        -\sin\delta & -\cos\delta\\
        \cos\delta & -\sin\delta
    \end{bmatrix}
    \begin{bmatrix}
        t\\
        n
    \end{bmatrix}.
\end{equation}

 Using the transformation matrix and the projection in the normal direction of the equation of motion \eqref{eq:eom}, one can express the turn rate as
\begin{equation}
    \omega=-\frac{V_n}{H\dot{\varphi}}=-\frac{n_av^a|\nabla V|}{\dot{\varphi}H}=\frac{\cos\delta}{c}, \label{eq:omegadelta}
\end{equation}
where\footnote{The parameter $c$ was introduced in \cite{Anguelova:2022foz}, it will not play a role in what follows.} $c=\frac{\dot{\varphi}H}{|\nabla V|}$. Similarly, dividing the projection in the tangential direction of the equation of motion \eqref{eq:eom} by $H\dot{\varphi}$, $\eta_{\parallel}$ can be written in terms of $\delta$ as
\begin{equation}
    -\eta_{\parallel}=3+\frac{V_t}{H\dot{\varphi}}=3+\frac{t_av^a|\nabla V|}{H\dot{\varphi}}=3-\frac{\sin\delta}{c}.\label{eq:etadelta}
\end{equation}
From  equations \eqref{eq:omegadelta} and \eqref{eq:etadelta} one finds that
the angle $\delta$ in the transformation matrix is related to $\eta_{\parallel}$ and $\omega$ by \textit{an exact relation}
\begin{equation}
    \label{eq:omegaeta}
   3+ \eta_{\parallel}=\omega\tan\delta.
\end{equation}
We will use this equation later to relate $\omega$ and $\delta$ in the slow-roll regime. In this case $\eta_{\parallel}\ll 1$ therefore $\omega \tan\delta \sim 3$. For very large turn rates $|\omega|\gg1$ it follows that $\delta \sim 3/\omega \ll 1$. Moderate turn rates $|\omega|\sim\mathcal{O}(1)$ allow for larger values of $\delta$ around $\pi/4$. Notice that both regimes are in principle consistent with the rapid-turn condition $\omega^2\gg {\cal O}($\rm slow-roll$)$ that leads to attractor behaviour.

On the other hand,  $\delta = 0$ (a trajectory with $V=$ const., orthogonal to $\nabla V$) requires $\eta_{\parallel} = -3$, not consistent with slow-roll. This is the ultra-slow-roll regime \cite{Inoue:2001zt,Kinney:2005vj} which is beyond the scope of this paper and left for future work. Similarly, we do not discuss the case with $\nabla V =0$ in what follows.

\section{Consistency conditions for slow-roll rapid-turn two-field inflation}

\label{sec:cc}
In this section we discuss the consistency conditions for sustained slow roll and rapid turn inflation following \cite{Bjorkmo:2019fls, Anguelova:2022foz}. We further derive a generalized consistency condition for rapid-turn inflation in the potential rather than the kinematical basis, without any assumptions on the turning rate of the inflationary trajectory. The potential basis allows for scanning the parameter space for rapid-turn solutions, contrary to the kinematical basis where $t$ and $n$ (and thus $V_{tt}$ etc.) are not known without solving the full equations of motion.
\subsection{The rapid-turn consistency condition}
From the equations of motion \eqref{eq:eom} it follows \cite{Hetz:2016ics, Achucarro:2010da, Chakraborty:2019dfh}
\begin{align}
\begin{split}\label{eq:VttVtn1}
    V_{tt}&=\Omega^2+H^2(3\varepsilon_H-3\eta_{\parallel}-\xi_{\parallel}),\\
    V_{tn}&=-\Omega H(3-\varepsilon_H+\nu_H+2\eta_{\parallel}),
    \end{split}
\end{align}
where $V_{tt}$, $V_{tn}$ are projections of the covariant Hesse matrix along tangential and normal vectors defined as $V_{\alpha\beta}=\alpha_a\beta_b\nabla^a\nabla^bV(\phi)$. The parameter $\nu_H$ quantifies the change of $\omega$ per e-fold and $\xi_{\parallel}$ is the third slow-roll parameter, defined as
\begin{equation}
    \label{eq:def_xi_nu}
  \nu_H=\frac{\dot{\omega}}{H\omega},  \quad \xi_{\parallel}=\frac{\dddot{\varphi}}{H^2\dot{\varphi}}.
\end{equation}
We assume both $\nu_H, \xi_{\parallel}\ll 1$.
Dividing equation \eqref{eq:VttVtn1} by $H^2$ and using $V=H^2(3-\varepsilon_H)$ gives
\begin{align}
\begin{split}
    \frac{V_{tt}}{V}&=\frac{1}{3-\varepsilon_H}\left(\omega^2+3\varepsilon_H-3\eta_{\parallel}-\xi_{\parallel}\right), \label{eq:VttVtn}\\
    \frac{V_{tn}}{V}&=-\frac{\omega}{3-\varepsilon_H}(3-\varepsilon_H+\nu_H+2\eta_{\parallel}).
    \end{split}
\end{align}
Before we proceed to the consistency condition, notice that  
we can substitute for $\omega$ and obtain the following \textit{exact} relation
\begin{equation}
    \label{eq:consistency_exact}
    \frac{V_{tt}}{V}=\frac{3-\varepsilon_H}{(3-\varepsilon_H+\nu_H+2\eta_{\parallel})^2}\left(\frac{V_{tn}}{V}\right)^2 + \frac{\varepsilon_H(3-\eta_{\parallel})-\xi_{\parallel}}{3-\varepsilon_H}.
\end{equation}
This will be important in section \ref{kappa}. Neglecting slow-roll corrections gives \textit{the rapid-turn condition in the kinematical basis}
\begin{equation}\label{eq:ALconsistencycond}
    \frac{V_{tt}}{V}=\frac{1}{3}\left(\frac{V_{tn}}{V}\right)^2.
\end{equation}

First, we aim to re-express the consistency condition \eqref{eq:ALconsistencycond} in the potential basis.
Using the basis transformation from the kinematical basis to the potential basis \eqref{eq:basistransformation} one finds
\begin{align}
\begin{split}
    V_{tt}&=V_{vv}\sin^2\delta-2V_{vw}\sin\delta\cos\delta+V_{ww}\cos^2\delta,\label{eq:transformVtt}\\
    V_{tn}&=(V_{vv}-V_{ww})\cos\delta\sin\delta+V_{vw}\left(\sin^2\delta-\cos^2\delta\right).
    \end{split}
\end{align}
To express the angle $\delta$ through functions in the potential basis we use \eqref{eq:VttVtn}, assuming $\omega^2\gg {\cal O}(\rm slow-roll)$, and equation $3+ \eta_{\parallel}=\omega\tan\delta$ \eqref{eq:omegaeta} to get a quadratic equation for $\tan \delta$
\begin{equation}
  \frac{V_{tt}}{V} \tan^2 \delta + \frac{V_{tn}}{V} \tan\delta = 0 + ... , \label{eq:Vttdelta}
\end{equation}
where ... indicates terms of order slow roll that have been neglected. Plugging \eqref{eq:transformVtt} into \eqref{eq:Vttdelta}
leads to 
\begin{equation}
    \frac{V_{vv}}{V} \tan^2 \delta - \frac{V_{vw}}{V} \tan\delta = 0 + ...
\end{equation}
Of the two solutions,  we reject $\tan \delta = 0$ (that was introduced to obtain the quadratic equation),  and are left with 
\begin{equation}
\label{eq:tandelta}
\tan\delta = \frac{V_{vw}}{V_{vv}} + ...
\end{equation}
Hence, we find
\begin{equation}
    \label{eq:delta}
    \delta=\arctan\frac{V_{vw}}{V_{vv}}.
 \end{equation}
 The exact expression for $\tan \delta$ and its dependence on slow-roll parameters is derived in Appendix \ref{appendix:delta}.

We now re-express the rapid-turn condition \eqref{eq:ALconsistencycond} in the potential basis using \eqref{eq:transformVtt} and \eqref{eq:tandelta}. 
This
leads to \textit{the rapid turn condition} for any value of turning rate
\begin{equation}
    \label{eq:genConsCond}
    \frac{V_{ww}}{V}=3+3\left(\frac{V_{vv}}{V_{vw}}\right)^2+\frac{V_{vv}}{V}\left(\frac{V_{vw}}{V_{vv}}\right)^2.
\end{equation}

In the regime of very small $\delta$, the condition found previously in reference \cite{Anguelova:2022foz} is recovered. One finds 
\begin{equation}
\label{eq:anguelova_result}
\begin{split}
    \csc^2\delta&\approx\frac{1}{\delta^2}+\mathcal{O}(1),\\
    \frac{V_{vw}}{V_{vv}}=\tan\delta&\approx\delta+\mathcal{O}(\delta^3),
\end{split}
\end{equation}
hence we get the result from \cite{Anguelova:2022foz}
\begin{equation}
  \frac{V_{ww}}{V}\approx 3\left(\frac{V_{vv}}{V_{vw}}\right)^2. \label{eq:ALcondition}
\end{equation}

\subsection {The rapid turn condition is not sufficient to guarantee slow roll}
\label{kappa}

The rapid turn condition
\eqref{eq:genConsCond} was derived assuming slow roll, but it is also compatible with potentially large values of the slow roll parameters. To see this,  consider again equation 
\eqref{eq:consistency_exact} with the slow roll parameters satisfying
the following relations
\begin{subequations}
\label{eq:al_regime}
\begin{align}
    \label{eq:al_regime1}
    3(3-\varepsilon_H)=(3-\varepsilon_H+\nu_H+2\eta_{\parallel})^2,\\
    \label{eq:al_regime2}
    \varepsilon_H(3-\eta_{\parallel})-\xi_{\parallel}=0.
\end{align}
\end{subequations}
Then the previous conditions become exact
\begin{equation}
\label{eq:omegaVtt}
\begin{split}
    \frac{V_{tt}}{V}&=\frac{\omega^2}{3-\varepsilon_H}\\
    \frac{V_{tn}}{V}&=-\sqrt{\frac{3}{3-\varepsilon_H}}\omega
\end{split}
\quad \Rightarrow \quad \frac{V_{tt}}{V}=\frac{1}{3}\left(\frac{V_{tn}}{V}\right)^2 .
\end{equation}

However,  the necessary assumptions in equation (\ref{eq:al_regime}) are more general than the slow-roll regime: with two equations and four unknowns one can find solutions of (\ref{eq:al_regime}) with large $\varepsilon_H$ and $\eta_{\parallel}$. The slow-roll regime ($\varepsilon_H=\eta_{\parallel}=\nu_H=\xi_{\parallel}=0$) is only one of many possible solutions of equations (\ref{eq:al_regime}). One has therefore to keep in mind that slow-roll inflation is not guaranteed
when the condition \eqref{eq:ALconsistencycond} holds, 
we need an additional condition. This can be done in a number of ways, since all we require is an expression that reduces to $\epsilon_H$ on the trajectories that satisfy the rapid-turn condition $\omega^2\gg {\cal O}(\rm slow-roll)$. 

In particular, when $\omega^2\gg {\cal O}(\rm slow-roll)$ we can express the slow-roll parameter using \eqref{eq:epsilonV}, \eqref{eq:VtviaepsH} and \eqref{eq:epsilon_V_exact} as
\begin{equation} \label{eq:epsilon_VttVt}
    \varepsilon_H=\frac{3\left(\varepsilon_V-\frac{1}{2}\left(\frac{V_t}{V} \right)^2 \right)}{\varepsilon_V+\frac{V_{tt}}{V}-\frac{1}{2}\left(\frac{V_t}{V}\right)^2} \ ,
\end{equation}
where we replaced $\omega$ using 
$V_{tt}/V \approx \omega^2/(3-\epsilon_H)$
and
\begin{align}
     \frac{1}{2}\left(\frac{V_t}{V}\right)^2&=\varepsilon_V\, \frac{V_{vw}^2}{V_{vv}^2+V_{vw}^2},\label{eq:VtoverV}\\
    V_{tt}&=\frac{V_{ww}V_{vv}^2-V_{vv}V_{vw}^2}{V_{vw}^2+V_{vv}^2}.  \label{eq:vtt_potentialBasis}
\end{align}
Equations \eqref{eq:VtoverV} and \eqref{eq:vtt_potentialBasis} simplify the relation \eqref{eq:epsilon_VttVt} to
\begin{equation}
    \frac{3\, \varepsilon_V}{\varepsilon_V+
\left( \frac{V_{ww}}{V} - \frac{V_{vw}^2}{V V_{vv}} \right) }\ll 1 , \label{eq:epsilonHCondition}
\end{equation}
with $\varepsilon_V$ given by \eqref{eq:epsilonV}.

The equation \eqref{eq:epsilonHCondition} provides \textit{a slow-roll condition} on the rapid-turn attractor. 
It is written entirely in terms of (derivatives of) the scalar potential and the geometry of the scalar manifold and can be used directly to verify if the slow-roll regime $\varepsilon_H\ll1$ holds when the consistency condition \eqref{eq:genConsCond} is satisfied. We discuss the numerical implementation of the consistency condition and the slow-roll condition in the next section.

\section{Numerical implementation} \label{sec:numerics}
In this Section we describe the numerical implementation of the consistency conditions and verify our results with three models of rapid-turn inflation. We further compare our findings with the result from reference \cite{Anguelova:2022foz}. 
\subsection{The \texttt{inflatox} software package}

\begin{figure}[h]
    \centering
    \includegraphics{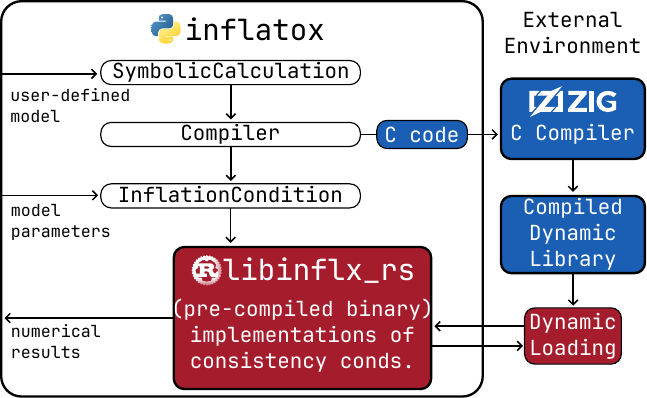}
    \caption{Schematic representation of the \texttt{inflatox} software package implementation. }
    \label{fig:inflatox_design}
\end{figure}

The \texttt{inflatox} software package provides a numerical implementation of the consistency conditions \eqref{eq:genConsCond} and \eqref{eq:epsilonHCondition}, as well as the result from \cite{Anguelova:2022foz} in equation \eqref{eq:ALcondition}. It was designed with parameter searches in mind: a user inputs an effective two-field model and explores the parameter space of a given model in search of slow-roll rapid-turn trajectories. Both numerical performance and ease of use were important factors in the design of the software package, which is written in Python (user-facing code) and Rust (numerical code). The package architecture is shown in figure \ref{fig:inflatox_design}.

Besides the potential, field-space metric and components of the Hesse matrix, the \texttt{inflatox} also provides the numerical implementations for the following derived quantities:
\begin{enumerate}
    \item The consistency condition \eqref{eq:genConsCond}.
    \item The leading order consistency condition \eqref{eq:ALcondition} from \cite{Anguelova:2022foz}.
    \item The potential slow-roll parameter $\varepsilon_V$ \eqref{eq:epsilonV}.
    \item The slow-roll condition 
    \eqref{eq:epsilonHCondition}.
    \item The angle $|\delta|\in[0,\pi/2]$ \eqref{eq:delta}.
    \item The turn-rate $\omega$ \eqref{eq:omegaVtt} using \eqref{eq:vtt_potentialBasis}-\eqref{eq:epsilonHCondition}
     for $V_{tt}$ and $\varepsilon_H$.
\end{enumerate}
The consistency condition and its leading order version are returned in the following format:
\begin{equation}
    \label{eq:out}
    \rm{out}=\frac{\left|\rm{lhs}-\rm{rhs}\right|}{|\rm{lhs}|+|\rm{rhs}|}\in[0,1]
\end{equation}
Where $\rm{lhs}$ and $\rm{rhs}$ are the left- and right-hand-side of the consistency condition. Thus, $\rm{out}=0$ indicates that the consistency condition is perfectly satisfied, whereas $\rm{out}=1$ indicates that it is maximally violated. 

\begin{figure}
    \centering
    \includegraphics[width=.7\textwidth]{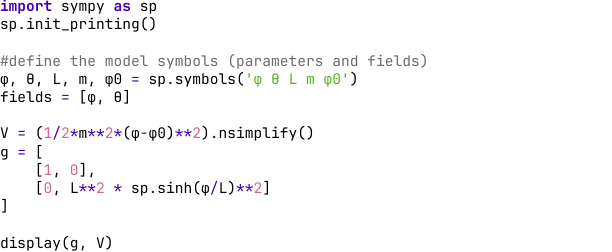}
    \caption{Example showing the definition (input) of a two-field inflation model using sympy.}
    \label{fig:inputinflatox}
\end{figure}

Users can input their models in \texttt{inflatox} using the \texttt{sympy} \cite{Meurer:2017yhf} symbolic mathematics package. In sympy, mathematical symbols are declared by the user which can then be combined in equations using regular Python syntax. The example of the input model is shown in figure \ref{fig:inputinflatox}.

Using sympy, the user supplies inflatox with a scalar potential and field space metric, after which it can symbolically calculate the components of the projected Hesse matrix $V_{\alpha\beta}=\alpha^a\beta^b\nabla_a\nabla_bV(\phi)$. The potential basis $\{v,w\}$ can be calculated by inflatox: $v$ is derived directly from the user-supplied potential (as per equation (\ref{eq:potential_basis})) and $w$ is calculated using the covariant Gramm-Schmidt procedure. The user has to specify a starting point for the vector $w$, as is usual for the \textsc{gs} procedure. The orientation of the $\{v,w\}$ basis depends on the user-supplied starting point, which may in turn change the signs of $\delta$ and $\omega$.

Inflatox can translate the symbolic expressions for the scalar potential and metric into executable C code. This C-code is compiled using the \texttt{zig-cc} compiler \cite{zig}, which is packaged with inflatox as a dependency. The compiled model is then handed over to the (pre-compiled) Rust backend of inflatox: a pre-compiled binary called \texttt{libinflx\_rs}. The Rust code can directly call the functions in the compiled model binary to calculate the consistency condition(s) or perform a parameter sweep.

All this underlying complexity is completely hidden from the user. The user only ever has to interact with the top-level Python package, which exposes all the functionality of the underlying Rust implementation. Inflatox uses Python's C-API to communicate between Python and Rust. 

Inflatox has been published to the Python package repository \texttt{PyPi}. It can be downloaded directly from \href{https://pypi.org/project/inflatox/}{https://pypi.org/project/inflatox/} or by installing it through \texttt{pip} (\texttt{pip install inflatox}). The source code of the project can be found on Github at\\ \href{https://github.com/smups/inflatox/}{https://github.com/smups/inflatox/}.

\subsection{Example models}\label{sec:models}
In this section, we present results obtained using \texttt{inflatox} for two-field models of inflation with known slow-roll rapid-turn solutions. We compare the \texttt{inflatox} results with numerical solutions to the equations of motion found using PyTransport and Mathematica. We refer the reader to 
the original papers for details on the models.

\subsubsection{Angular Inflation}
The `Angular inflation' model \cite{Christodoulidis:2018qdw} provides a simple example of sustained rapid-turn slow-roll inflation that is a dynamical attractor along the boundary of the Poincaré disc\footnote{Note that, because of the hyperbolic geometry, the boundary is at an infinite proper distance from the trajectory.}. The field-space metric and the potential of the model are given by
\begin{align}
    \label{eq:angular_metric}
    G_{ab}&=\frac{6\alpha}{(1-\chi^2-\phi^2)^2}\boldsymbol{1}_{ab},\\
     \label{eq:angular_potential}
    V(\chi,\phi)&=\frac{\alpha}{2}\left(m_{\phi}^2\phi^2+m_{\chi}^2\chi^2\right).
\end{align}
The metric (\ref{eq:angular_metric}) describes the Poincaré disk, which is a hyperbolic space of constant negative field-space curvature $ \mathbb{R}=-\frac{3}{4\alpha}$. For numerical computations the parameters listed in table \ref{tab:angular_parameters} were used.

\begin{table}[h]
\centering
\begin{tabular}{r|c|c|c|c|c|c|c|c|c|c}
\textbf{Parameter} & $\alpha$ & $m_{\phi}$ & $m_{\chi}$ \\ \hline
\textbf{Value} & $1/600$ & $2\times10^{-5}$ & $3m_{\phi}$
\end{tabular}
\caption{Values used for the parameters of the angular inflation model.}
\label{tab:angular_parameters}
\end{table}

\subsubsection{EGNO model}
The Ellis-García-Nanopoulos-Olive (EGNO) model \cite{Ellis:2014gxa} is a scale-free supergravity model with a single dynamical superfield $\Phi=r+i\theta$, the real and imaginary parts of which are two dynamical scalar fields during inflation. This model was further discussed in the context of rapid-turn models of inflation in \cite{Aragam:2021scu}. From the Kähler metric and superpotential, the following field-space metric and potential arise
\begin{align}
    \label{eq:egno_metric}
    G_{ab}(r)&=-3\alpha\boldsymbol{1}_{ab}\left[\frac{12c(2r-1)^2}{c(2r-1)^{4}-2r}-\left(\frac{4c\left(2r-1\right)^{3} - 1}{c(2r-1)^{4}-2r}\right)^2\right],\\
    \label{eq:egno_potential}
    V(r,\theta)&=\frac{6m^2}{a^2}r^3\left(\theta^2+(a-r)^2\middle)\middle(2r-c(1-2r)^4\right).
\end{align}
In the numerical section of this work, the parameter values listed in the table \ref{tab:egno_parameters} were used.

\begin{table}[h]
\centering
\begin{tabular}{r|c|c|c|c|c|c|c|c|c|c}
\textbf{Parameter} & $\alpha$ & $m$ & $a$ & $c$ \\ \hline
\textbf{Value} & $1$ & $10^{-3}$ & $1/2$ & $10^3$
\end{tabular}
\caption{Values used for the parameters of the EGNO supergravity model.}
\label{tab:egno_parameters}
\end{table}

\subsubsection{D5-brane model}
The last model we consider is the D5-brane first introduced in \cite{Kenton:2014gma} and later studied in the context of multifield inflation in \cite{Chakraborty:2019dfh}. Being an effective field theory derived from a string theory model, the expressions for the potential and field space metric are very involved and are given by
\begin{equation}
    \mathcal{H}(r)=\frac{\pi Ng_sl_s^4}{12u^4}\left(\frac{2}{\rho^2}-2\ln\left\{\frac{1}{\rho^2}+1\right\}\right),
    \quad \rho=\frac{r}{u}
    \label{eq:d5_first}
\end{equation}
\begin{equation}
    G_{ab}(r)=\frac{4\pi\mu_5}{g_s}\sqrt{\frac{\mathcal{H}}{9}(r^2+3u^2)^2+(\pi l_s^2q)^2}\times\text{diag}\left\{\frac{r^2+6u^2}{r^2+9u^2},\frac{1}{6}(r^2+6u^2)\right\},
\end{equation}
\begin{equation}
\label{eq:stringtheory_potential}
\begin{split}
    \overline{\Phi}_-(r)=&\frac{5}{72}\left[81\rho^2(9\rho^2-2)\rho^2+162\ln\{9\rho^2+9\}-9-160\ln10\right],\\
    \Phi_h(r,\theta_2)=&a_0\left[\frac{2}{\rho^2}-2\ln\left\{\frac{1}{\rho^2}+1\right\}\right] + 2a_1\left[6+\frac{1}{\rho^2}-(4+6\rho^2)\ln\left\{\frac{1}{\rho^2}+1\right\}\right]\cos\theta_2\\
    +& \frac{b_1}{2}(2+3\rho^2)\cos\theta_2,
\end{split}
\end{equation}
\begin{equation}
    V(r,\theta_2)=V_0+\frac{4\pi pT_5}{\mathcal{H}}\left(\left[\frac{\mathcal{H}}{9}(r^2+3u^2)^2+(\pi l_s^2q)^2\right]^{-1/2}-l_s^2\pi qg_s\right)+4\pi^2l_s^2pqT_5g_s(\overline{\Phi}_-+\Phi_h).
    \label{eq:d5_potential}
\end{equation}
Equations \eqref{eq:d5_first}-\eqref{eq:d5_potential} contain in total ten parameters. Table \ref{tab:d5_parameters} shows the parameters used for numerical computations.

\begin{table}[h]
\centering
\begin{tabular}{r|c|c|c|c|c|c|c|c|c|c}
\textbf{Parameter} & $V_0$ & $N$ & $g_s$ & $l_s$ & $u$ & $q$ & $p$ & $a_0$ & $a_1$ & $b_1$ \\ \hline
\textbf{Value} & $-1.17\times10^{-8}$ & $1000$ & $0.01$ & $501.961$ & $50l_s$ & $1$ & $5$ & $0.001$ & $a_0/2$ & $a_0$
\end{tabular}
\caption{Values used for the parameters of the D5-brane model.}
\label{tab:d5_parameters}
\end{table}

\subsection{Results }
\begin{figure}
\begin{center}
\includegraphics[width=1\textwidth]{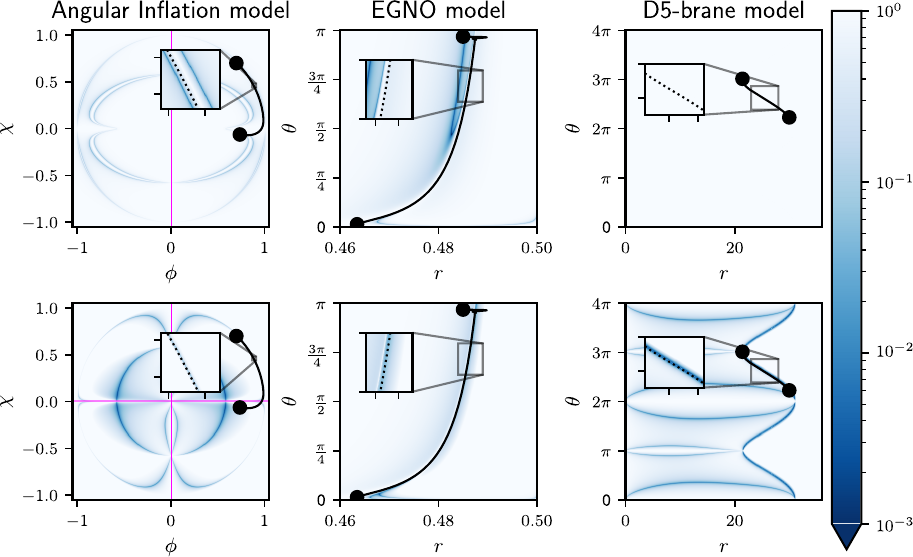}
\end{center}
\caption[]{Comparison of the condition \eqref{eq:ALcondition} \cite{Anguelova:2022foz} (top panel) and the generalized rapid-turn condition \eqref{eq:genConsCond} (lower panel) for angular inflation, EGNO and D5-brane inflation models (from left to right respectively). The validity of the consistency conditions is indicated by the normalised difference between the right- and left-hand-sides of the equation, as defined in \eqref{eq:out}. The blue color indicates regions where the consistency condition holds while white regions show where it is violated. The black curves show exact numerical solutions of the equations of motion. On the insets numerical trajectories are indicated as dotted black curves.
}
\label{fig:conscond}
\end{figure}

In this Section we show the result of numerical implementation of consistency condition \eqref{eq:genConsCond} and the condition \eqref{eq:ALcondition} from \cite{Anguelova:2022foz} for the Angular inflation, EGNO and D5-brane inflation models described in Section \ref{sec:models}. These models are known in the literature to have rapid-turn attractors with turning rates $\omega\sim {\cal O}(1)$ for Angular inflation and EGNO models, and $\omega\sim {\cal O}(10)$ for D5-brane inflation (for the parameters in Section \ref{sec:models}).
In figure \ref{fig:conscond} we show the result for both rapid-turn consistency conditions. We check the validity of the consistency conditions at each point of the field-space parameter space, indicated by the normalized difference between the right- and left-hand-sides of the equation, as defined in \eqref{eq:out}. The blue areas on the panels indicate that the rapid-turn attractor condition is met, while the white areas indicate its violation. Black solid curves show the exact trajectories found numerically for each model.

On the top panel of figure \ref{fig:conscond}
we show the results for the condition \eqref{eq:ALcondition} \cite{Anguelova:2022foz}.
The condition \eqref{eq:ALcondition} works with good accuracy for the Angular inflation model. However, it does not identify correctly the EGNO solution and fails to predict the D5-brane attractor, as shown in more detail on the insets of figure \ref{fig:conscond}.

As a comparison, on the bottom panel of figure \ref{fig:conscond} we show the consistency condition \eqref{eq:genConsCond} introduced in this work. The condition \eqref{eq:genConsCond} identifies the exact rapid-turn solutions with excellent accuracy - blue regions where the consistency condition indicates the rapid-turn attractor are exactly on top of the black curves showing exact solutions.

\begin{figure}
\begin{center}
\includegraphics{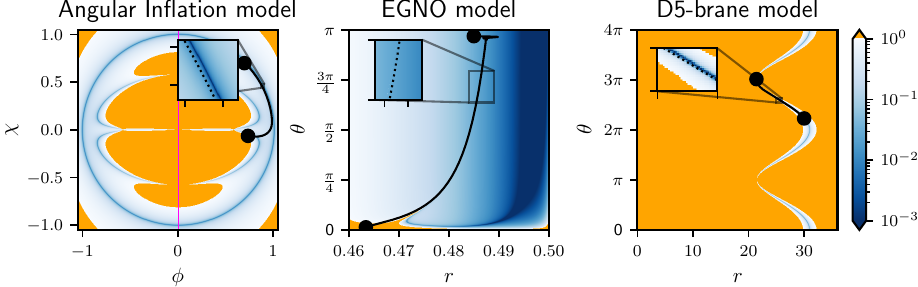}
\end{center}
\caption[]{ The slow-roll condition \eqref{eq:epsilonHCondition} for Angular inflation, EGNO and D5-brane inflation models (from left to right respectively). The yellow color indicates regions in field-space where the slow-roll condition does not hold. The blue color indicates small values of the slow-roll condition. The black curves (dotted black on the insets) show exact numerical solutions of the equations of motion.
}
\label{fig:epsilon}
\end{figure}

However, we see that both conditions \eqref{eq:ALcondition} and \eqref{eq:genConsCond} are satisfied also in 
regions of parameter space far from the actual numerical trajectories. As discussed in Section \ref{kappa} this is because the rapid-turn attractor conditions do not guarantee the slow-roll condition is satisfied. To ensure the rapid-turn trajectories are also slow-roll, it is necessary to check that condition  \eqref{eq:epsilonHCondition} holds. In figure \ref{fig:epsilon} we show the result of the numerical implementation of slow-roll condition \eqref{eq:epsilonHCondition} for Angular inflation, EGNO and D5-brane models. The yellow areas on the panels indicate that the slow-roll condition is larger than one and hence in those regions solutions to the rapid-turn condition do not give slow-roll trajectories. To check that the rapid-turn attractors are also slow roll, one has to check the overlap of
the rapid-turn condition \eqref{eq:genConsCond} and the slow-roll condition \eqref{eq:epsilonHCondition}, i.e. super-impose the results from figures \ref{fig:conscond} and \ref{fig:epsilon}. We see from figure \ref{fig:epsilon} that the slow-roll condition is satisfied only in small regions around exact numerical trajectories. This excludes all other regions where the rapid-turn condition holds away from the exact trajectories in figure \ref{fig:conscond} and demonstrates that the only slow-roll rapid-turn solutions are given by the known attractors in these models. To sum up, \textit{both} conditions \eqref{eq:genConsCond} and  \eqref{eq:epsilonHCondition} are needed to identify rapid-turn slow-roll trajectories.

\section{Summary and outlook}\label{sec:conclusion}

Multi-field inflation models with non-geodesic motion and field-space curvature are receiving increasing attention because of their interesting, potentially detectable, predictions for primordial spectra and because of their arguably more natural connection to particle physics models beyond the standard model. 

A particularly interesting class of models, known as slow-roll rapid-turn inflation, has been shown to be a (perturbative) attractor \cite{Bjorkmo:2019fls}. If sustained for enough e-folds, these inflation models could give primordial spectra compatible with current observations on the largest scales, yet potentially distinguishable from single-field inflation by the next generation of cosmological observations. However, these models are difficult to find in practice and only a handful of examples are known in the literature. Recently an analytical rapid-turn consistency condition was derived in \cite{Anguelova:2022foz} for very large turning rates.

In this work we have extended this analysis and presented a method and the Python package \texttt{inflatox} for identifying slow-roll rapid-turn inflationary trajectories in two-field inflation for any values of the turn rate $\omega$ that are consistent with the attractor condition $\omega^2\gg {\cal O}(\rm slow-roll)$. Besides the rapid-turn equation, a second condition is needed to ensure slow-roll.

An important open question is whether the rapid-turn slow-roll conditions can be sustained for enough e-folds of inflation to produce the observed primordial spectra on CMB scales. Furthermore, our numerical implementation revealed regions in field-space where the turn rate is nearly constant and the rapid-turn condition holds, however slow-roll is violated. Therefore, our results may also facilitate the identification of ultra-slow-roll and other non-slow-roll multi-field behaviours and their systematic study. It will be interesting to explore the nature and potential applications of solutions that violate slow-roll. The extension of this method to models with more than two fields is very interesting and challenging and is left for future work.

\section*{Acknowledgments}

We thank L. Anguelova and P. Christodoulidis for useful comments and discussions, as well as S. Paban, R. Rosati, I. Zavala and the participants of the Simons Center workshop ``Multifield Cosmology: Inflation, Dark Energy and More" where this work was presented.
We are grateful to the Simons Center for Geometry and Physics for their hospitality and support. The work of O.I. was supported by the European Union's Horizon 2020
research and innovation program under the Marie Skłodowska-Curie grant
agreement No.~101106874.
O.I.\ is grateful to the University of Leiden for hospitality, where parts of this work have been completed. Nordita was sponsored by Nordforsk.

\paragraph{Note added.} While completing this manuscript, reference \cite{Anguelova:2024akm} appeared which also revises and extends the results 
of \cite{Anguelova:2022foz, Anguelova:2023mpe} and has some overlap with this work. Our conclusions agree and our discussions are complementary. 
We thank Lilia Anguelova for discussions and comments.

\appendix
\section{Full derivation of transformation angle}\label{appendix:delta}

In this section we derive the exact expression for $\tan \delta$ including all slow-roll parameters.
We start with equation \eqref{eq:VttVtn} and write it in the form
\begin{align}
\begin{split}
    \frac{V_{tt}}{V}&=\frac{\omega^2+\tilde{\alpha}}{3-\varepsilon_H},\\
    \frac{V_{tn}}{V}&=-\omega(1+\tilde{\beta}),\label{eqap:VttVtn}
    \end{split}
\end{align}
with $\tilde{\alpha}=3\varepsilon_H-3\eta_{\parallel}-\xi_{\parallel}$ and $\tilde{\beta}=\frac{\nu_H+2\eta_{\parallel}}{3-\varepsilon_H}$. Similarly as in Section \ref{sec:cc} we multiply equations \eqref{eqap:VttVtn} by $\tan^2 \delta$ and $\tan \delta$  to get a quadratic equation for $\tan \delta$. We find the exact relation
\begin{equation}
  \frac{V_{tt}}{V} \tan^2 \delta + \frac{V_{tn}}{V} \tan\delta =\frac{3+\eta_{\parallel}}{3-\varepsilon_H}\left(\varepsilon_H -\eta_{\parallel}-\nu_H\right)+\frac{\tilde{\alpha}}{3-\varepsilon_H}\tan ^2\delta,
\end{equation}
where we used equation $3+ \eta_{\parallel}=\omega\tan\delta$ \eqref{eq:omegaeta}. Using the exact relation 
\begin{equation}
    \frac{V_{tt}}{V} \tan^2 \delta + \frac{V_{tn}}{V} \tan\delta = \frac{V_{vv}}{V} \tan^2 \delta - \frac{V_{vw}}{V} \tan\delta,
\end{equation}
we find the quadratic equation for $\tan \delta$
\begin{equation}
   \left( \frac{V_{vv}}{V}-\frac{\tilde{\alpha}}{3-\varepsilon_H} \right)\tan^2 \delta - \frac{V_{vw}}{V} \tan\delta-\frac{3+\eta_{\parallel}}{3-\varepsilon_H}\left(\varepsilon_H -\eta_{\parallel}-\nu_H\right)=0.
\end{equation}
Solving for $\tan \delta$ gives the exact expression
\begin{equation}\label{eq:tanDeltaExact}
  \tan \delta=\frac{\frac{V_{vw}}{V}\pm \sqrt{ \left(\frac{V_{vw}}{V} \right)^2 +4\left(\frac{V_{vv}}{V} -\frac{\tilde{\alpha}}{3-\varepsilon_H}\right) \frac{3+\eta_{\parallel}}{3-\varepsilon_H}\left(\varepsilon_H -\eta_{\parallel}-\nu_H\right)       }  }
  {2\left( \frac{V_{vv}}{V} -\frac{\tilde{\alpha}}{3-\varepsilon_H} \right)},
\end{equation}
 provided $V_{vv}/ V \neq \tilde\alpha / (3-\varepsilon_H)$, otherwise
\begin{equation}
    \tan\delta = - \left(\frac{V}{V_{vw}}\right)
    \frac{3+\eta_{\parallel}}{3-\varepsilon_H}\left(\varepsilon_H -\eta_{\parallel}-\nu_H\right) \ . 
\end{equation}

The equation \eqref{eq:tanDeltaExact} reduces to the equation \eqref{eq:tandelta}
\begin{equation}
    \tan \delta\approx \frac{V_{vw}}{V_{vv}}
\end{equation}
when, for instance,
\begin{gather}
    \frac{\tilde{\alpha}}{3-\varepsilon_H}\ll \frac{V_{vv}}{V}\quad {\rm and} \quad 4 \, \frac{3+\eta_{\parallel}}{3-\varepsilon_H}\left(\varepsilon_H -\eta_{\parallel}-\nu_H\right) \frac{V_{vv}}{V} \ll \left( \frac{V_{vw}}{V}\right)^2.
\end{gather}
More generally, the exact relation 
\begin{equation}
    \tan \delta= \frac{V_{vw}}{V_{vv}}
\end{equation}
holds when $V_{vv}/ V \neq \tilde\alpha / (3-\varepsilon_H)$ and
\begin{equation}
   \left( \frac{V_{vv}}{V}-\frac{\tilde{\alpha}}{3-\varepsilon_H} \right)
   \left(\frac{V}{V_{vw}}\right)^2
   (3+\eta_{\parallel}) \left(\varepsilon_H -\eta_{\parallel}-\nu_H\right)
   - \frac{\tilde\alpha^2}{3-\varepsilon_H}\frac{V^2}{V_{vv}^2}
   +\tilde\alpha\frac{V}{V_{vv}}
    =0.
\end{equation}

\bibliographystyle{JHEP}
\bibliography{refs.bib}

\end{document}